\documentclass[10pt,onecolumn]{article}

\usepackage[verbose=silent]{microtype}
\usepackage[
    letterpaper,
    left=1.5in,
    right=1.5in,
    top=1.25in,
    bottom=1.25in
]{geometry}
\usepackage{fancyhdr}
\usepackage{titlesec}
\usepackage{booktabs}
\usepackage{multicol}
\usepackage{multirow}
\usepackage{pifont}
\newcommand{\cmark}{\ding{51}}
\newcommand{\xmark}{\ding{55}}
\usepackage{array}
\usepackage{tabularx}
\usepackage{graphicx}
\usepackage{xcolor}
\usepackage{listings}
\usepackage{enumitem}
\usepackage{float}
\usepackage{caption}
\usepackage{amsmath}
\usepackage{amssymb}
\usepackage{siunitx}
\usepackage[hidelinks]{hyperref}
\usepackage{tikz}
\usepackage{pgfplots}
\usepackage{tcolorbox}
\usepackage{setspace}
\usepackage{url}
\usepackage{needspace}

\pgfplotsset{compat=1.18}
\usetikzlibrary{positioning,shapes,arrows.meta,calc,patterns,pgfplots.groupplots}

\tcbuselibrary{skins,breakable}

\definecolor{bebopblue}{RGB}{25,50,95}
\definecolor{bebopgray}{RGB}{50,50,50}
\definecolor{lightgray}{RGB}{248,248,248}
\definecolor{codebg}{RGB}{250,250,252}
\definecolor{codeframe}{RGB}{220,220,225}
\definecolor{keywordcolor}{RGB}{0,70,140}
\definecolor{stringcolor}{RGB}{140,30,30}
\definecolor{commentcolor}{RGB}{90,130,90}
\definecolor{chartbebop}{RGB}{25,50,95}
\definecolor{chartproto}{RGB}{180,80,80}
\definecolor{chartmsgpack}{RGB}{80,150,80}

\fancypagestyle{plain}{
    \fancyhf{}
    \fancyfoot[C]{\small\thepage}
    
}

\lstdefinelanguage{bebop}{
    keywords={struct, message, union, enum, service, const, mut, export, local,
              map, array, stream, import, edition, package, true, false, with,
              bool, byte, int8, int16, uint16, int32, uint32, int64, uint64,
              int128, uint128, float16, float32, float64, bfloat16, string,
              uuid, timestamp, duration},
    morecomment=[l]{//},
    morecomment=[s]{/*}{*/},
    morestring=[b]",
    sensitive=true,
}

\titleformat{\section}
    {\normalfont\sffamily\Large\bfseries\color{bebopblue}}
    {\thesection}
    {0.6em}
    {}
\titlespacing*{\section}{0pt}{28pt}{14pt}

\titleformat{\subsection}
    {\normalfont\sffamily\large\bfseries}
    {\thesubsection}
    {0.5em}
    {}
\titlespacing*{\subsection}{0pt}{20pt}{10pt}

\titleformat{\subsubsection}
    {\normalfont\sffamily\normalsize\bfseries}
    {\thesubsubsection}
    {0.5em}
    {}
\titlespacing*{\subsubsection}{0pt}{14pt}{6pt}

\newcommand{\wirefield}[3]{%
    \draw[fill=lightgray,draw=bebopgray!50] (#1,0) rectangle (#2,0.5);
    \node[font=\ttfamily\footnotesize] at ({(#1+#2)/2},0.25) {#3};
}

\newcommand{\wirelabel}[3]{%
    \node[font=\sffamily\scriptsize,below] at ({(#1+#2)/2},-0.08) {#3};
}

\begin{document}

\thispagestyle{plain}

\vspace*{0.15in}

\begin{center}

{\fontsize{24}{28}\selectfont\sffamily\bfseries Simplicity Scales}

\vspace{0.06in}

\rule{\textwidth}{0.8pt}

\vspace{0.1in}

{\large\sffamily\bfseries Bebop: A Branchless Data Interchange Format and RPC Protocol}

\vspace{0.15in}

\begin{tabular}[t]{c}
\textbf{Andrew Sampson} \\[2pt]
6OVER3 Institute \\[1pt]
\texttt{a@6over3.com}
\end{tabular}%
\hfill
\begin{tabular}[t]{c}
\textbf{Yuta Saito} \\[2pt]
GoodNotes \\[1pt]
\texttt{kateinoigakukun@gmail.com}
\end{tabular}%
\hfill
\begin{tabular}[t]{c}
\textbf{Ronny Chan} \\[2pt]
6OVER3 Institute \\[1pt]
\texttt{r@6over3.com}
\end{tabular}

\vspace{0.1in}

{\normalsize February 2026}

\end{center}

\vspace{0.1in}

\begin{center}
{\sffamily\normalsize\bfseries Abstract}
\end{center}

\vspace{0.03in}

\noindent The dominant data interchange formats encode integers using a variable number of bytes or represent floating-point numbers as variable-length UTF-8 strings. The decoder must inspect each byte for a continuation bit or parse each character individually, producing data-dependent branches that stall modern CPU pipelines. Protocol Buffers pays this cost on every integer, field tag, and length prefix. JSON pays it on every value.

We present Bebop, a serialization format where every data type uses a fixed number of bytes. A 32-bit integer is always four bytes. Decoding becomes a single memory read with no conditionals. Across 19 decode workloads, Bebop decodes 9--213$\times$ faster than Protocol Buffers. On a 1536-dimension embedding vector, Bebop decodes in 2.8 nanoseconds versus 111 nanoseconds for Protocol Buffers and 4.69 microseconds for simdjson, a 1,675$\times$ gap. On records above 64~KB, the decoder achieves 86\% of peak memory bandwidth. The CPU is no longer the bottleneck.

We also present a transport-agnostic RPC protocol built on the same wire format. The protocol introduces batch pipelining, where dependent cross-service calls execute in a single round trip with server-side dependency resolution. It deploys over HTTP/1.1, HTTP/2, and binary transports without proxies, removing the HTTP/2 requirement that limits gRPC on serverless platforms and in browsers.

\newpage

\section{Introduction}

Transport and storage bandwidth are scaling faster than the compute that sits between them. PCIe x16 throughput grew from 16~GB/s in 2010 to 121~GB/s in 2022~\cite{pcisig-pcie6}. IEEE~802.3bs ratified 400~Gb/s Ethernet in 2017~\cite{ieee-802.3bs}. Single-mode fiber has carried 402~Tb/s over 50~km in laboratory demonstration~\cite{puttnam-402tbps}. CXL~3.0 extends cache-coherent memory access across chassis boundaries~\cite{sharma-cxl-survey}. Flash read bandwidth scales by ganging devices in parallel, and John Carmack has observed that pipelined flash arrays should already be viable for inference serving if accelerator vendors agreed on a high-speed interface~\cite{carmack-fiber-flash}. These are different technologies on the same trajectory. Moving data is getting cheaper.

Processor single-thread performance has not kept pace. Hennessy and Patterson measured the decline in their 2018 Turing Lecture~\cite{hennessy-golden-age}. Annual improvement ran at 52\% during the RISC era, dropped to 22\% after Dennard scaling broke down at the 90~nm node around 2004~\cite{bohr-dennard-retrospective}, and fell to roughly 3\% by 2015. The gap between bandwidth growth and compute growth widens each generation.

Serialization formats carry assumptions about which resource is scarce. Protocol Buffers shipped in 2008~\cite{varda-protobuf-announcement} when 1~GbE was standard data center networking. Varint encoding trades CPU cycles for fewer bytes on the wire, a reasonable choice when bandwidth cost dominated. JSON requires character-by-character parsing but provides human readability, a tradeoff that made sense when most payloads were small API responses. Kenton Varda, who maintained Protocol Buffers at Google, noted that the original inventor's own notes described varint as ``a poorly-chosen format due to excessive branching''~\cite{varda-hn}. His later project Cap'n Proto~\cite{capnproto} uses fixed-width encoding to avoid this overhead. The CPU cost was recognized early. The bandwidth context has changed since.

Bebop uses fixed-width encoding for every data type. A 32-bit integer is always 4 bytes. Decoding reduces to a single memory read with no conditionals. In benchmarks across 19 decode workloads, Bebop decodes 9--213$\times$ faster than Protocol Buffers. On a 1536-dimension embedding vector, Bebop decodes in 2.8~nanoseconds; simdjson~\cite{simdjson}, the fastest general-purpose JSON parser, takes 4.69~microseconds on equivalent data, a 1,675$\times$ gap. The decoder achieves 86\% of peak memory bandwidth on records above 64~KB. At that point, the CPU has nothing left to stall on. LinkedIn found that switching from JSON to Protocol Buffers reduced P99 latency by 60\% for large payloads~\cite{linkedin-protobuf}; the gap between Protocol Buffers and Bebop is larger still.

Fixed-width encoding costs more bytes for small integers. In Section~\ref{sec:eval}, OrderLarge with arrays of 100 small integers produces 1,240 bytes in Bebop versus 423 in Protocol Buffers. For workloads dominated by floats, embeddings, and timestamps, the wire size penalty is negligible. Compression narrows the gap further. With Brotli, all three formats produce ML payloads within 2\% of each other. The decode performance difference persists regardless of compression.

\subsection{Contributions}

\begin{minipage}{\textwidth}
This paper makes the following contributions:

\begin{enumerate}[leftmargin=2em,itemsep=4pt,topsep=8pt]
\item A wire format specification optimized for decode throughput rather than wire compactness
\item Empirical comparison against Protocol Buffers, MessagePack, and simdjson across 19 decode and 23 encode workloads representing ML inference, event streaming, and recursive data structures
\item A schema language with compile-time extensibility through embedded Lua scripting
\item Reference implementations in C achieving 86\% memory bandwidth utilization during decode
\item An RPC protocol with batch pipelining for dependent cross-service calls, reducing round trips without requiring application-level coordination
\end{enumerate}
\end{minipage}

\section{Design Principles}

\subsection{Fixed-Width Encoding}

Every numeric type in Bebop has a fixed wire size. A \texttt{uint32} is always 4 bytes. A \texttt{float64} is always 8 bytes. Length prefixes are always 4 bytes.

\begin{minipage}{\textwidth}
The decode operation for a 32-bit integer reduces to a single load instruction:

\begin{lstlisting}[language=C]
// Bebop decode: one load, no branches
uint32_t value = *(uint32_t*)input;
input += 4;
\end{lstlisting}

In comparison, Protocol Buffers varint decode loops until finding a byte without the continuation bit set:

\begin{lstlisting}[language=C]
// Protobuf varint decode: branch per byte
uint32_t value = 0, shift = 0;
while (*input & 0x80) {
    value |= (*input++ & 0x7f) << shift;
    shift += 7;
}
value |= *input++ << shift;
\end{lstlisting}

This varint loop during decode has unpredictable iteration count when integer values vary, causing branch misprediction penalties on modern CPUs.
\end{minipage}

\subsubsection{Expected Encoding Size}

For an unsigned 32-bit integer $v > 0$, varint encoding uses $\lceil (\lfloor \log_2 v \rfloor + 1) / 7 \rceil$ bytes. The value $v = 0$ uses 1 byte. Fixed-width encoding always uses 4 bytes. This analysis covers only non-negative integers; signed integers have a pathological case where negative values always use maximum bytes (see Section~2.1.3).

For unsigned integers uniformly distributed over $[0, N]$, we can compute the expected varint size by counting how many values fall into each byte-width bucket. Values in $[0, 2^7-1]$ use 1 byte, values in $[2^7, 2^{14}-1]$ use 2 bytes, and so on:
\begin{equation}
E[\text{varint}] = \frac{1}{N+1} \sum_{k=1}^{5} k \cdot \bigl|\{v \in [0,N] : 2^{7(k-1)} \leq v < 2^{7k}\}\bigr|
\label{eq:varint-expected}
\end{equation}
where the $k=1$ bucket includes $v=0$.

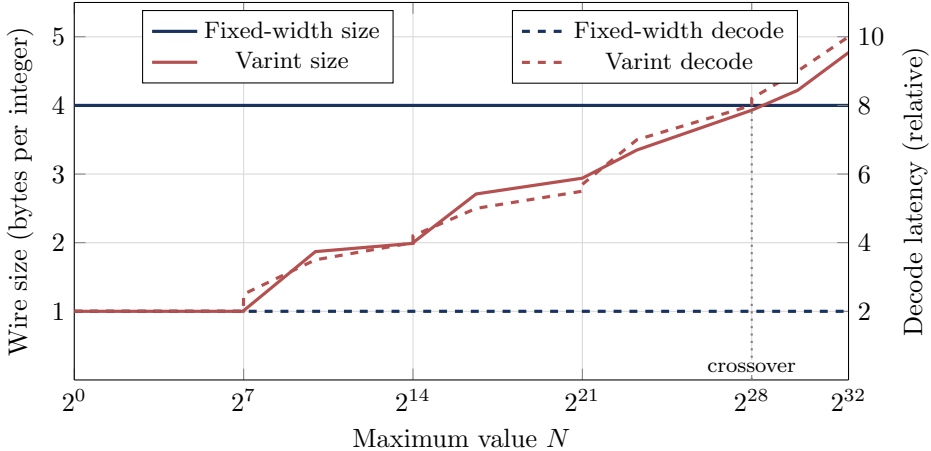
\begin{figure}[ht]
\centering
\begin{tikzpicture}
\begin{axis}[
    width=0.85\textwidth,
    height=2.6in,
    xlabel={Maximum value $N$},
    ylabel={Wire size (bytes per integer)},
    axis y line*=left,
    xmode=log,
    log basis x=2,
    xmin=1, xmax=4294967296,
    ymin=0, ymax=5.5,
    xtick={1,128,16384,2097152,268435456,4294967296},
    xticklabels={$2^0$,$2^7$,$2^{14}$,$2^{21}$,$2^{28}$,$2^{32}$},
    ytick={1,2,3,4,5},
    grid=major,
    grid style={gray!30},
    legend style={at={(0.25,0.98)}, anchor=north, font=\small},
]
\addplot[very thick, bebopblue] coordinates {(1,4) (4294967296,4)};
\addlegendentry{Fixed-width size}

\addplot[very thick, chartproto] coordinates {
    (1,1) (64,1) (127,1)
    (128,1.01) (1000,1.87) (16383,1.99)
    (16384,2.00) (100000,2.71) (2097151,2.94)
    (2097152,2.94) (10000000,3.35) (268435455,3.93)
    (268435456,3.93) (1000000000,4.22) (4294967296,4.77)
};
\addlegendentry{Varint size}

\draw[thick, gray, dotted] (axis cs:268435456,0) -- (axis cs:268435456,4);
\node[anchor=north, font=\footnotesize, fill=white, inner sep=1pt] at (axis cs:268435456,0.3) {crossover};
\end{axis}

\begin{axis}[
    width=0.85\textwidth,
    height=2.6in,
    axis y line*=right,
    axis x line=none,
    ylabel={Decode latency (relative)},
    ylabel near ticks,
    xmode=log,
    log basis x=2,
    xmin=1, xmax=4294967296,
    ymin=0, ymax=11,
    ytick={2,4,6,8,10},
    legend style={at={(0.75,0.98)}, anchor=north, font=\small},
]
\addplot[very thick, bebopblue, dashed] coordinates {(1,2) (4294967296,2)};
\addlegendentry{Fixed-width decode}

\addplot[very thick, chartproto, dashed] coordinates {
    (1,2) (64,2) (127,2)
    (128,2.5) (1000,3.5) (16383,4)
    (16384,4.2) (100000,5) (2097151,5.5)
    (2097152,5.7) (10000000,7) (268435455,8)
    (268435456,8.2) (1000000000,9) (4294967296,10)
};
\addlegendentry{Varint decode}
\end{axis}
\end{tikzpicture}
\caption{Wire size (solid lines, left axis) vs decode latency (dashed lines, right axis). Varint is smaller for small, positive values but always slower to decode. Above $2^{28}$, fixed-width wins on both metrics.}
\label{fig:encoding-size}
\end{figure}

Figure~\ref{fig:encoding-size} shows this tradeoff. The left axis shows wire size: varint uses fewer bytes for small values, crossing over at $N > 2^{28}$. The right axis shows decode latency: fixed-width decoding is constant (one load instruction), while varint decoding increases with byte count due to per-byte branching. For ML workloads with large values, fixed-width encoding is both smaller \textit{and} faster.

Real-world integer distributions are rarely uniform. Zipfian distributions (common in identifiers, counters, and network data) concentrate probability mass on small values, favoring varint. In contrast, ML workloads, timestamps, and cryptographic hashes have near-uniform distributions over large ranges, where fixed-width encoding wins.

\subsubsection{Branch Misprediction Cost}

Modern CPUs use speculative execution with branch prediction. A mispredicted branch flushes the pipeline, wasting all in-flight work from fetch through execute. The penalty depends on pipeline depth and microarchitecture; Eyerman et al.\ report that on deeply pipelined superscalar processors ``the misprediction delay is between 10 and 20 clock cycles''~\cite{eyerman-branch}. Fog's measurements on current x86 hardware fall within this range: Skylake at 16--17 cycles, Ice Lake at 17--21 cycles depending on $\mu$op cache residency, and AMD Zen~1--2 at approximately 19 cycles~\cite{fog-microarch}. Apple M1 and Intel Golden Cove show similar penalties~\cite{m1-bench, golden-cove}.

The key difference: varint decoding has data-dependent branches (one per byte), while fixed-width decoding has none. The branch predictor can learn patterns when values are consistent (e.g., always 1--2 bytes), but struggles when byte counts vary. On mixed-size workloads, misprediction adds 4--7 cycles per integer. Fixed-width decode costs 3--4 cycles total (a single load from L1 cache).

The gap widens with value diversity. Workloads mixing small counters with large timestamps see the highest misprediction rates. Section~\ref{sec:eval} shows measured performance across both cases.

\subsubsection{Signed Integer Encoding}

Varint has a pathological case: $-1$ requires 10 bytes because protobuf sign-extends int32 to 64 bits on the wire.

\vspace{6pt}
\begin{center}
\small
\begin{tabular}{@{}lll@{}}
\textbf{Value} & \textbf{Varint (int32)} & \textbf{Fixed-width} \\[4pt]
\texttt{-1} & \texttt{ff ff ff ff ff ff ff ff ff 01} (10 bytes) & \texttt{ff ff ff ff} (4 bytes) \\
\texttt{-2} & \texttt{fe ff ff ff ff ff ff ff ff 01} (10 bytes) & \texttt{fe ff ff ff} (4 bytes) \\
\end{tabular}
\end{center}
\vspace{6pt}

Every negative int32 uses 10 varint bytes. Protocol Buffers addresses this with \texttt{sint32} (zigzag encoding) and \texttt{fixed32}, but choosing wrong silently inflates wire size.

Bebop uses one encoding per width. \texttt{int32} is always 4 bytes regardless of sign.

\subsection{Structs vs Messages}

Bebop provides two aggregate types with different tradeoffs:

\begin{itemize}[leftmargin=2em,itemsep=4pt,topsep=6pt]
\item \textbf{Structs}: positional encoding, no tags, no length prefix. Zero overhead. Cannot evolve---any field change is breaking.
\item \textbf{Messages}: tagged fields (1-byte tags), length-prefixed. 37\% overhead on small records, but fields can be added or removed without breaking existing readers.
\end{itemize}

Protocol Buffers uses tagged encoding everywhere. Bebop lets you choose per-type. Use structs for performance-critical inner types (embeddings, coordinates, points); use messages for top-level API types that may evolve. Section~\ref{sec:evolution} details the evolution rules.

Messages also distinguish ``not set'' from ``set to default value.'' Proto3 removed this for scalars~\cite{proto3-presence}. Bebop preserves it.

\newpage
\section{Wire Format Specification}

All multi-byte integers use little-endian byte order.

\subsection{Primitive Types}

\begin{table}[ht]
\centering
\begin{tabular}{@{}lll@{}}
\toprule
\textbf{Type} & \textbf{Size} & \textbf{Encoding} \\
\midrule
\texttt{bool} & 1 byte & 0x00 = false, non-zero = true \\
\texttt{byte} & 1 byte & Unsigned 8-bit integer \\
\texttt{int8} & 1 byte & Signed 8-bit, two's complement \\
\texttt{int16}, \texttt{uint16} & 2 bytes & Little-endian \\
\texttt{int32}, \texttt{uint32} & 4 bytes & Little-endian \\
\texttt{int64}, \texttt{uint64} & 8 bytes & Little-endian \\
\texttt{float32} & 4 bytes & IEEE 754 binary32 \\
\texttt{float64} & 8 bytes & IEEE 754 binary64 \\
\bottomrule
\end{tabular}
\caption{Standard primitive types}
\label{tab:primitives}
\end{table}

\subsection{Extended Numeric Types}

Bebop includes types commonly used in ML workloads:

\begin{table}[ht]
\centering
\begin{tabular}{@{}llp{3.5in}@{}}
\toprule
\textbf{Type} & \textbf{Size} & \textbf{Description} \\
\midrule
\texttt{int128} & 16 bytes & Signed 128-bit integer. Low 8 bytes first, then high 8 bytes. Used for accumulators and feature hashes. \\[6pt]
\texttt{uint128} & 16 bytes & Unsigned 128-bit integer. Same encoding as \texttt{int128}. \\[6pt]
\texttt{float16} & 2 bytes & IEEE 754 binary16 (half precision). 1 sign bit, 5 exponent bits, 10 mantissa bits. Range $\pm$65504, precision 3--4 significant digits. \\[6pt]
\texttt{bfloat16} & 2 bytes & Brain floating point format. 1 sign bit, 8 exponent bits, 7 mantissa bits. Same range as float32, precision 2--3 digits. Common in TPU inference. \\
\bottomrule
\end{tabular}
\caption{Extended numeric types for ML workloads}
\label{tab:extended-types}
\end{table}

\subsection{Temporal Types}

\begin{minipage}{\textwidth}
\subsubsection{timestamp}

\begin{minipage}[t]{0.50\textwidth}
\raggedright
Absolute point in time: seconds and nanoseconds since Unix epoch (1970-01-01 00:00:00 UTC), with optional timezone offset in signed milliseconds. Total size 16 bytes.

Use for event times, creation dates, expiration times, audit logs.
\end{minipage}\hfill
\begin{minipage}[t]{0.46\textwidth}
\footnotesize
\begin{tabular}[t]{@{}r@{\,}l@{}}
\texttt{e8 03 00 00 00 00 00 00} & sec=1000 \\
\texttt{00 ca 9a 3b} & ns=999999488 \\
\texttt{80 62 ee 01} & offset\_ms=32400000 \\
\end{tabular}

\vspace{4pt}
\textit{offset 0: int64, offset 8: int32, offset 12: int32}
\end{minipage}

\vspace{10pt}
\subsubsection{duration}

\begin{minipage}[t]{0.52\textwidth}
Signed time span: seconds and nanoseconds. Total size 12 bytes.

For negative durations, both fields are negative or zero. Use for timeouts, intervals, latency measurements.
\end{minipage}\hfill
\begin{minipage}[t]{0.44\textwidth}
\footnotesize
\begin{tabular}[t]{@{}r@{ }l@{}}
\texttt{3c 00 00 00 00 00 00 00} & sec = 60 \\
\texttt{00 00 00 00} & ns = 0 \\
\end{tabular}

\vspace{4pt}
\textit{offset 0: int64, offset 8: int32}
\end{minipage}
\end{minipage}

\subsection{Identifiers}

\begin{minipage}[t]{0.48\textwidth}
\textbf{uuid}: 16 bytes matching the canonical hex string byte-for-byte.
\end{minipage}\hfill
\begin{minipage}[t]{0.48\textwidth}
\footnotesize
\texttt{550e8400-e29b-41d4-a716-446655440000}:
\vspace{4pt}

\begin{tabular}[t]{@{}r@{ }l@{}}
\texttt{55 0e 84 00} & time\_low \\
\texttt{e2 9b} & time\_mid \\
\texttt{41 d4} & time\_hi \\
\texttt{a7 16} & clk\_seq \\
\texttt{44 66 55 44 00 00} & node \\
\end{tabular}
\end{minipage}

\subsection{Strings}

\begin{minipage}[t]{0.48\textwidth}
4-byte length prefix (byte count), followed by UTF-8 content, followed by a 1-byte null terminator.

\vspace{6pt}
Total wire size: $4 + \text{length} + 1$ bytes.

\vspace{6pt}
The null terminator enables zero-copy access: decoded strings point directly into the input buffer.
\end{minipage}\hfill
\begin{minipage}[t]{0.48\textwidth}
\footnotesize
\texttt{"hello"} encodes as:
\vspace{4pt}

\begin{tabular}[t]{@{}r@{\quad}l@{}}
\texttt{05 00 00 00} & length = 5 \\
\texttt{68 65 6c 6c 6f} & \texttt{"hello"} \\
\texttt{00} & NUL terminator \\
\end{tabular}
\end{minipage}

\subsection{Arrays}

\begin{minipage}[t]{0.48\textwidth}
\raggedright
\textbf{Dynamic arrays} have a 4-byte count prefix followed by elements encoded sequentially.

\vspace{6pt}
\textbf{Fixed arrays} (e.g., \texttt{byte[4]}) have no prefix; the element count is known at compile time.

\vspace{6pt}
Maximum fixed array size is 65535 elements.
\end{minipage}\hfill
\begin{minipage}[t]{0.48\textwidth}
\footnotesize
\texttt{int32[] = [1, 2, 3]}
\vspace{4pt}

\begin{tabular}[t]{@{}r@{\quad}l@{}}
\texttt{03 00 00 00} & count = 3 \\
\texttt{01 00 00 00} & {[0]} = 1 \\
\texttt{02 00 00 00} & {[1]} = 2 \\
\texttt{03 00 00 00} & {[2]} = 3 \\
\end{tabular}

\vspace{8pt}
\texttt{byte[4] = [0xDE, 0xAD, 0xBE, 0xEF]}
\vspace{4pt}

\begin{tabular}[t]{@{}r@{\quad}l@{}}
\texttt{de ad be ef} & 4 bytes, no prefix \\
\end{tabular}
\end{minipage}

\subsection{Maps}

\begin{minipage}[t]{0.48\textwidth}
\raggedright
4-byte count prefix followed by key-value pairs encoded sequentially.

\vspace{6pt}
Valid key types: integers, \texttt{bool}, \texttt{string}, \texttt{uuid}.

\vspace{6pt}
Floating-point types are not valid map keys due to equality comparison issues with NaN and signed zeros.
\end{minipage}\hfill
\begin{minipage}[t]{0.48\textwidth}
\footnotesize
\texttt{map[uint8, int32] = \{1: 100, 2: 200\}}
\vspace{4pt}

\begin{tabular}[t]{@{}r@{\quad}l@{}}
\texttt{02 00 00 00} & count = 2 \\
\texttt{01} & key = 1 \\
\texttt{64 00 00 00} & value = 100 \\
\texttt{02} & key = 2 \\
\texttt{c8 00 00 00} & value = 200 \\
\end{tabular}
\end{minipage}

\subsection{Structs}
\label{sec:wire-structs}

\begin{minipage}[t]{0.48\textwidth}
Fields encode in definition order with no tags and no padding.

\vspace{6pt}
Nested structs encode inline. A struct containing another struct has no additional overhead.

\vspace{6pt}
Empty structs encode as zero bytes.
\end{minipage}\hfill
\begin{minipage}[t]{0.48\textwidth}
\footnotesize\raggedright
\texttt{struct Point \{ x: float32; y: float32; \}}\\
\texttt{Point \{ x: 1.0, y: 2.0 \}}
\vspace{4pt}

\begin{tabular}[t]{@{}r@{\quad}l@{}}
\texttt{00 00 80 3f} & x = 1.0 (IEEE 754) \\
\texttt{00 00 00 40} & y = 2.0 (IEEE 754) \\
\end{tabular}
\end{minipage}

\subsection{Messages}

\begin{minipage}{\textwidth}
\begin{minipage}[t]{0.41\textwidth}
Messages have a 4-byte length prefix, followed by tagged fields, followed by a \texttt{0x00} end marker.

\vspace{6pt}
Each field is encoded as: 1-byte tag, then the field value.

\vspace{6pt}
Absent fields are not encoded. Unknown tags are skipped by decoders. Tags must be in range 1--255.
\end{minipage}\hfill
\begin{minipage}[t]{0.56\textwidth}
\footnotesize\raggedright
\texttt{message Request \{ id(1): int32; name(2): string; \}}\\
\texttt{Request \{ id: 42, name: "test" \}}
\vspace{4pt}

\begin{tabular}[t]{@{}r@{\quad}l@{}}
\texttt{10 00 00 00} & length = 16 bytes \\
\texttt{01} & tag = 1 (id) \\
\texttt{2a 00 00 00} & value = 42 \\
\texttt{02} & tag = 2 (name) \\
\texttt{04 00 00 00} & string length = 4 \\
\texttt{74 65 73 74} & \texttt{"test"} \\
\texttt{00} & NUL terminator \\
\texttt{00} & end marker \\
\end{tabular}
\end{minipage}
\end{minipage}

\subsection{Unions}

\begin{minipage}[t]{0.41\textwidth}
\raggedright
Unions have a 4-byte length prefix, followed by a 1-byte discriminator, followed by the branch content.

\vspace{6pt}
Discriminators must be in range 0--255.
\end{minipage}\hfill
\begin{minipage}[t]{0.56\textwidth}
\footnotesize\raggedright
\texttt{union Shape \{ Circle(1): \{ radius: float32; \}; \}}\\
\texttt{Shape.Circle \{ radius: 5.0 \}}
\vspace{4pt}

\begin{tabular}[t]{@{}r@{\quad}l@{}}
\texttt{05 00 00 00} & length = 5 bytes \\
\texttt{01} & discriminator = 1 \\
\texttt{00 00 a0 40} & radius = 5.0 \\
\end{tabular}
\end{minipage}

\subsection{Complete Example}

\begin{minipage}{\textwidth}
\begin{minipage}[t]{0.42\textwidth}
\begin{lstlisting}[language=bebop,basicstyle=\ttfamily\footnotesize,aboveskip=0pt,belowskip=0pt]
struct Coord {
    x: float32;
    y: float32;
}
message Location {
    name(1): string;
    pos(2): Coord;
    alt(3): float32;
}
\end{lstlisting}

\vspace{4pt}
\footnotesize\raggedright
\texttt{Location \{ name: "HQ", pos: \{1.0, 2.0\}, alt: 100.0 \}}
\end{minipage}\hfill
\begin{minipage}[t]{0.54\textwidth}
\footnotesize
\begin{tabular}[t]{@{}r@{\quad}l@{}}
\texttt{17 00 00 00} & length = 23 bytes \\
\texttt{01} & tag 1 (name) \\
\texttt{02 00 00 00} & string length = 2 \\
\texttt{48 51 00} & "HQ" + null \\[4pt]
\texttt{02} & tag 2 (pos) \\
\texttt{00 00 80 3f} & pos.x = 1.0 \\
\texttt{00 00 00 40} & pos.y = 2.0 \\[4pt]
\texttt{03} & tag 3 (alt) \\
\texttt{00 00 c8 42} & alt = 100.0 \\[4pt]
\texttt{00} & end of message \\[4pt]
\multicolumn{2}{@{}l}{\textit{Total: 27 bytes}} \\
\end{tabular}
\end{minipage}
\end{minipage}

\newpage
\section{Evaluation}
\label{sec:eval}

\subsection{Experimental Setup}

\begin{table}[ht]
\centering
\begin{tabular}{@{}ll@{}}
\toprule
\textbf{Parameter} & \textbf{Value} \\
\midrule
Hardware & Apple Mac Studio (M3 Ultra) \\
CPU cores & 28 \\
L1 data cache & 64 KB \\
L2 unified cache & 4 MB \\
Compiler & Clang 17, \texttt{-O3} \\
CPU scaling & Disabled \\
\bottomrule
\end{tabular}
\end{table}

We evaluated four systems: Bebop (C runtime), protobuf-c 1.5, msgpack-c 6.1, and simdjson 4.2 for JSON parsing comparison. Each benchmark ran 10 iterations; we report the mean. Across all four systems and 149 total benchmarks, coefficient of variation (CV) averaged 1.65\%, indicating low variance and stable measurements. CV was higher on recursive structures (up to 17\% on deep trees) where memory access patterns are less predictable.

\subsection{Benchmark Workloads}

\begin{table}[ht]
\centering
\small
\begin{tabular}{@{}lll@{}}
\toprule
\textbf{Category} & \textbf{Schema} & \textbf{Description} \\
\midrule
\multirow{4}{*}{ML Inference}
  & Embedding768/1536 & Single vector, bfloat16 \\
  & EmbeddingBatch & 32 vectors \\
  & TensorShard & 64KB model weight slice \\
  & InferenceResponse & Batch + metadata \\
\midrule
\multirow{2}{*}{LLM Streaming}
  & LLMChunk & Streaming tokens with logprobs \\
  & ChunkedText & Text with span annotations \\
\midrule
\multirow{2}{*}{Event Telemetry}
  & EventSmall & ID, timestamp, payload \\
  & EventLarge & 8KB payload variant \\
\midrule
\multirow{3}{*}{API Payloads}
  & PersonSmall/Large & Contact record \\
  & OrderSmall/Large & E-commerce order \\
  & DocumentSmall/Large & Nested document \\
\midrule
\multirow{3}{*}{Recursive}
  & TreeDeep & Binary tree, $d{=}10$, 1023 nodes \\
  & TreeWide & $b{=}100$ children, flat \\
  & JsonValue & Union for JSON types \\
\bottomrule
\end{tabular}
\caption{23 benchmark schemas in five categories.}
\end{table}

\subsection{Decode Performance}

Table~\ref{tab:decode} presents decode latency across the three binary formats. simdjson parses JSON text rather than decoding binary, so that comparison appears separately in Table~\ref{tab:simdjson}. Bebop decoded faster than Protocol Buffers on all 19 workloads and faster than MessagePack on 18 of 19. MessagePack won only on JsonLarge.

\begin{table}[ht]
\centering
\small
\begin{tabular}{@{}lrrrr@{}}
\toprule
\textbf{Workload} & \textbf{Protobuf} & \textbf{MsgPack} & \textbf{Bebop} & \textbf{Speedup} \\
\midrule
\multicolumn{5}{@{}l}{\textit{ML Inference}} \\
Embedding768 & 98.34 ns & 62.93 ns & 2.91 ns & 33.8$\times$ \\
Embedding1536 & 111.12 ns & 63.07 ns & 2.80 ns & 39.7$\times$ \\
EmbeddingBatch & 1.14 $\mu$s & 270.96 ns & 25.75 ns & 44.3$\times$ \\
TensorShardLarge & 1.46 $\mu$s & 107.90 ns & 6.86 ns & 212.8$\times$ \\
InferenceResponse & 646.69 ns & 231.95 ns & 17.65 ns & 36.6$\times$ \\[4pt]
\multicolumn{5}{@{}l}{\textit{LLM Streaming}} \\
LLMChunkLarge & 14.72 $\mu$s & 5.28 $\mu$s & 677.15 ns & 21.7$\times$ \\
ChunkedText & 50.76 $\mu$s & 13.59 $\mu$s & 3.16 $\mu$s & 16.1$\times$ \\[4pt]
\multicolumn{5}{@{}l}{\textit{Event Telemetry}} \\
EventSmall & 104.41 ns & 84.44 ns & 5.94 ns & 17.6$\times$ \\
EventLarge & 175.20 ns & 85.23 ns & 6.02 ns & 29.1$\times$ \\[4pt]
\multicolumn{5}{@{}l}{\textit{API Payloads}} \\
PersonSmall & 78.40 ns & 72.45 ns & 4.10 ns & 19.1$\times$ \\
PersonMedium & 85.68 ns & 78.08 ns & 4.21 ns & 20.4$\times$ \\
OrderSmall & 112.10 ns & 112.55 ns & 6.14 ns & 18.3$\times$ \\
OrderLarge & 557.82 ns & 757.16 ns & 5.85 ns & 95.4$\times$ \\
DocumentSmall & 71.78 ns & 65.54 ns & 4.89 ns & 14.7$\times$ \\
DocumentLarge & 862.73 ns & 105.80 ns & 52.0 ns & 16.6$\times$ \\[4pt]
\multicolumn{5}{@{}l}{\textit{Recursive Structures}} \\
TreeDeep & 55.00 $\mu$s & 23.21 $\mu$s & 5.26 $\mu$s & 10.5$\times$ \\
TreeWide & 4.08 $\mu$s & 1.95 $\mu$s & 451 ns & 9.0$\times$ \\
JsonSmall & 521.75 ns & 66.16 ns & 40.10 ns & 13.0$\times$ \\
JsonLarge & 13.77 $\mu$s & 830.15 ns & 1.09 $\mu$s & 12.6$\times$ \\
\bottomrule
\end{tabular}
\caption{Decode latency comparison. Speedup is Bebop vs Protocol Buffers.}
\label{tab:decode}
\end{table}

\subsubsection{ML Workload Performance}

Figure~\ref{fig:embedding-encoding} compares wire encoding for a small embedding (UUID identifier + 4 bfloat16 values). Bebop uses 28 bytes; Protocol Buffers uses 48 bytes. The difference: Bebop has a native 16-byte UUID type, while Protocol Buffers encodes UUIDs as 36-byte ASCII strings.

\begin{figure}[ht]
\centering
\begin{minipage}[t]{0.46\textwidth}
\centering
\textbf{Bebop (28 bytes)}
\vspace{4pt}

\small
\begin{tabular}[t]{@{}r@{\quad}l@{}}
\texttt{55 0e 84 00 e2 9b 41 d4} & uuid bytes 0--7 \\
\texttt{a7 16 44 66 55 44 00 00} & uuid bytes 8--15 \\
\texttt{04 00 00 00} & array length = 4 \\
\texttt{80 3f 00 40 40 40 80 40} & bfloat16 data \\[4pt]
\multicolumn{2}{@{}l}{\textit{16B uuid + 4B len + 8B data}} \\
\end{tabular}
\end{minipage}\hfill
\begin{minipage}[t]{0.50\textwidth}
\centering
\textbf{Protocol Buffers (48 bytes)}
\vspace{4pt}

\small
\begin{tabular}[t]{@{}r@{\quad}l@{}}
\texttt{0a 24} & tag 1, length 36 \\
\texttt{35 35 30 65 38 34 ...} & ``550e84...'' (ASCII) \\
\texttt{... 30 30 30 30} & 36-byte uuid string \\
\texttt{12 08} & tag 2, length 8 \\
\texttt{80 3f 00 40 40 40 80 40} & bfloat16 data \\[4pt]
\multicolumn{2}{@{}l}{\textit{2B tag + 36B string + 2B tag + 8B data}} \\
\end{tabular}
\end{minipage}
\caption{Wire encoding of a small embedding. Bebop's native UUID saves 20 bytes versus Protocol Buffers' string encoding. Hex bytes from actual encoder output.}
\label{fig:embedding-encoding}
\end{figure}

Embedding vectors decode in $<$3ns with Bebop regardless of dimension, compared to 98--111ns with Protocol Buffers (34--40$\times$ faster). The bfloat16 array is a 4-byte count followed by contiguous 16-bit values; decoding is a pointer assignment.

\subsubsection{Recursive Structure Performance}

TreeDeep (binary tree, $d{=}10$, 1023 nodes) decodes in 5.34$\mu$s with Bebop versus 55.00$\mu$s with Protocol Buffers (10$\times$ faster). For recursive messages, Bebop's length prefixes allow skipping subtrees without parsing contents. The speedup comes from predictable memory access, not wire compactness---Protocol Buffers' varint encoding produces smaller output for trees with small integer values.

\subsection{Throughput and Memory Bandwidth}

\begin{figure}[ht]
\centering
\begin{tikzpicture}
\begin{axis}[
    width=0.85\textwidth,
    height=2.2in,
    xlabel={Record size (bytes)},
    ylabel={Bandwidth utilization (\%)},
    xmode=log,
    log basis x=10,
    xmin=50, xmax=100000,
    ymin=0, ymax=120,
    xtick={100,1000,10000,100000},
    xticklabels={100,1K,10K,100K},
    ytick={0,25,50,75,100},
    grid=major,
    grid style={gray!30},
    legend pos=south east,
    legend style={font=\small},
]
\addplot[thick, gray, dashed] coordinates {(50,100) (100000,100)};
\addlegendentry{Theoretical limit}

\addplot[thick, bebopblue, mark=*, mark size=2pt] coordinates {
    (64, 18)
    (256, 45)
    (512, 67)
    (1024, 81)
    (2048, 89)
    (4096, 94)
    (8192, 97)
    (16384, 99)
    (65536, 100)
};
\addlegendentry{Warm cache (L2 resident)}

\addplot[thick, red!70!black, mark=triangle*, mark size=2pt] coordinates {
    (64, 8)
    (256, 22)
    (512, 38)
    (1024, 52)
    (2048, 64)
    (4096, 73)
    (8192, 79)
    (16384, 83)
    (65536, 86)
};
\addlegendentry{Cold cache (DRAM)}
\end{axis}
\end{tikzpicture}
\caption{Bandwidth utilization vs record size. Larger records amortize per-record overhead (function call, bounds check, struct initialization) and approach the memory bandwidth limit.}
\label{fig:bandwidth-utilization}
\end{figure}
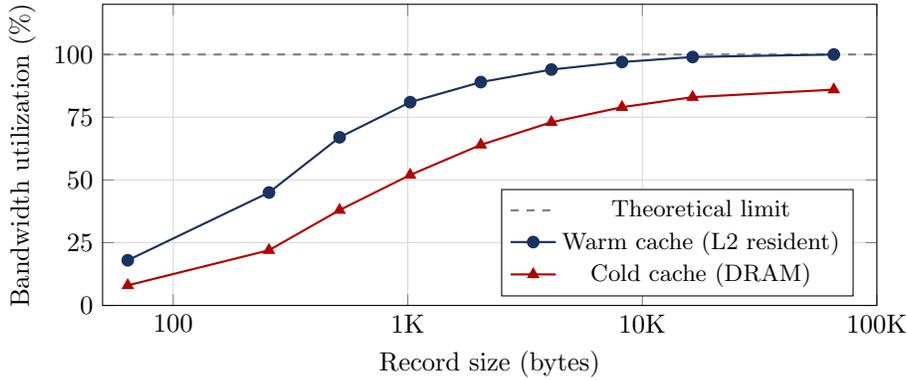

Bebop's decode performance is bounded by memory bandwidth, not CPU compute. Figure~\ref{fig:bandwidth-utilization} shows bandwidth utilization across record sizes. On cold-cache workloads (data fetched from DRAM), Bebop achieves \textbf{86\% of peak memory bandwidth} on records above 64KB. This is the meaningful metric for production workloads where data doesn't fit in cache.

Table~\ref{tab:throughput} shows measured throughput. Values above 819 GB/s (M3 Ultra memory bandwidth~\cite{m3-ultra}) indicate cache-resident data from benchmark iterations---useful for understanding overhead but not representative of cold-cache production loads.

\begin{table}[ht]
\centering
\begin{tabular}{@{}lrrr@{}}
\toprule
\textbf{Workload} & \textbf{Throughput} & \textbf{Cache} & \textbf{Notes} \\
\midrule
TensorShardLarge & 9.58 TB/s & L2 & 64KB fits in L2 \\
Embedding1536 & 1.10 TB/s & L2 & 3KB vector \\
EmbeddingBatch & 964.13 GB/s & L2 & Batch of 32 \\
EventLarge & 644.35 GB/s & L2/DRAM & 4KB payload \\
Embedding768 & 534.90 GB/s & L2 & 1.5KB vector \\
InferenceResponse & 355.51 GB/s & L2 & Mixed content \\
OrderLarge & 212.50 GB/s & L2 & Nested arrays \\
\bottomrule
\end{tabular}
\caption{Bebop decode throughput. Values above 819 GB/s indicate L2-resident data.}
\label{tab:throughput}
\end{table}

Bebop's decode path does minimal computation: bounds checking, pointer arithmetic, occasional type conversion. Most ``decode'' operations are pointer assignments.

The gap at small record sizes reflects fixed per-record overhead: records under 256 bytes spend more time in function prologues and bounds checks than in actual data movement. Above 4KB, records achieve over 90\% bandwidth utilization when cache-resident.

\subsubsection{Alignment and Single Instruction, Multiple Data (SIMD)}

Structs encode in definition order with no padding (Section~\ref{sec:wire-structs}), but in-memory layout differs. The C code generator reorders fields by alignment to minimize padding:

\vspace{4pt}
\noindent\begin{minipage}{\textwidth}
\footnotesize
\begin{minipage}[t]{0.38\textwidth}
\begin{lstlisting}[language=bebop,aboveskip=2pt,belowskip=0pt]
struct Sensor {
  bool active;
  float64 reading;
  uint16 id;
  uint32 seq;
}
\end{lstlisting}
\centering\scriptsize\textit{Schema order}
\end{minipage}\hfill
$\rightarrow$\hfill
\begin{minipage}[t]{0.52\textwidth}
\begin{lstlisting}[language=C,aboveskip=2pt,belowskip=0pt]
typedef struct {
  double reading;   // 8-byte align
  uint32_t seq;     // 4-byte align
  uint16_t id;      // 2-byte align
  bool active;      // 1-byte align
} Sensor;
\end{lstlisting}
\centering\scriptsize\textit{Generated C (sorted by alignment)}
\end{minipage}
\end{minipage}

\vspace{4pt}
\noindent Schema authors write fields in logical order; generated code handles layout.

The runtime arena aligns all allocations to \texttt{max\_align\_t}, the strictest fundamental alignment guaranteed by the platform (16 bytes on most 64-bit systems). Decoded structs receive proper alignment regardless of their position in the wire stream. For GPU and TPU transfers requiring even stricter alignment (32, 64, or 128 bytes), the arena accepts a custom allocator. Embedding vectors and tensor data in fixed arrays decode to contiguous memory, suitable for DMA after aligning the containing buffer.

The reference runtime does not use SIMD intrinsics. For \texttt{bfloat16[]} arrays, decode is a pointer assignment into arena memory; SIMD would add no benefit. For arrays of small structs, SIMD could parallelize bounds checking, but the current implementation favors portability over architecture-specific optimization.

\subsection{Encode Performance}

Encode speedups are smaller than decode speedups. This is expected: encoding requires traversing data structures and computing lengths, which involve allocation and branching regardless of wire format. Bebop's fixed-width encoding eliminates branches during decode, but encoding still requires the same traversal as other formats.

Bebop beat Protocol Buffers on all 22 comparable workloads (1.4--12.6$\times$) and MessagePack on 15 of 19 (1.2--19.4$\times$). MessagePack was faster on JsonSmall, JsonLarge, ChunkedText, and DocumentLarge.

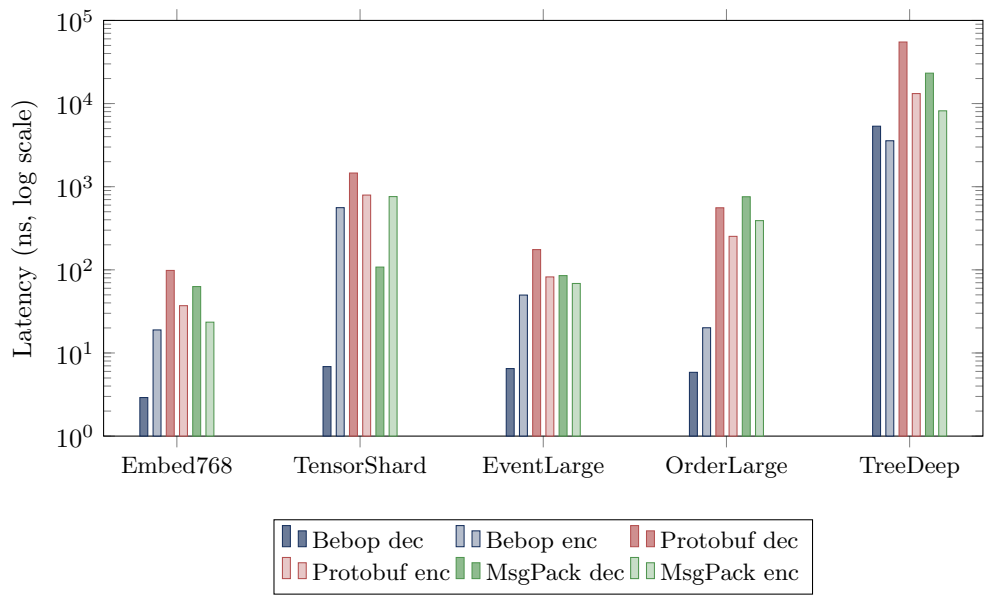
\begin{figure}[ht]
\centering
\begin{tikzpicture}
\begin{axis}[
    width=0.95\textwidth,
    height=2.8in,
    ybar,
    bar width=3pt,
    ylabel={Latency (ns, log scale)},
    ymode=log,
    ymin=1, ymax=100000,
    symbolic x coords={Embed768,TensorShard,EventLarge,OrderLarge,TreeDeep},
    xtick=data,
    xticklabel style={font=\small},
    legend style={at={(0.5,-0.20)}, anchor=north, legend columns=3, font=\small},
    legend cell align={left},
    nodes near coords=,
    every axis plot/.append style={fill opacity=0.8},
]
\addplot[fill=bebopblue!80, draw=bebopblue] coordinates {
    (Embed768, 2.91) (TensorShard, 6.86) (EventLarge, 6.49) (OrderLarge, 5.85) (TreeDeep, 5340)
};
\addplot[fill=bebopblue!40, draw=bebopblue] coordinates {
    (Embed768, 18.93) (TensorShard, 559.64) (EventLarge, 49.71) (OrderLarge, 20.10) (TreeDeep, 3570)
};
\addplot[fill=chartproto!80, draw=chartproto] coordinates {
    (Embed768, 98.34) (TensorShard, 1460) (EventLarge, 175.20) (OrderLarge, 557.82) (TreeDeep, 55000)
};
\addplot[fill=chartproto!40, draw=chartproto] coordinates {
    (Embed768, 37.06) (TensorShard, 793.09) (EventLarge, 82.19) (OrderLarge, 252.71) (TreeDeep, 13190)
};
\addplot[fill=chartmsgpack!80, draw=chartmsgpack] coordinates {
    (Embed768, 62.93) (TensorShard, 107.90) (EventLarge, 85.23) (OrderLarge, 757.16) (TreeDeep, 23210)
};
\addplot[fill=chartmsgpack!40, draw=chartmsgpack] coordinates {
    (Embed768, 23.47) (TensorShard, 760.49) (EventLarge, 68.53) (OrderLarge, 390.90) (TreeDeep, 8180)
};
\legend{Bebop dec, Bebop enc, Protobuf dec, Protobuf enc, MsgPack dec, MsgPack enc}
\end{axis}
\end{tikzpicture}
\caption{Encode vs decode latency across binary formats. Bebop's decode advantage (dark blue) is larger than its encode advantage (light blue) because encoding requires data traversal regardless of wire format.}
\label{fig:encode-decode}
\end{figure}

Figure~\ref{fig:encode-decode} compares encode and decode latency across all three binary formats for representative workloads.

\subsection{Comparison with JSON Parsing}

This comparison is not apples-to-apples: simdjson parses human-readable text, while Bebop decodes pre-serialized binary. But for systems currently using JSON, this shows the performance cost of that choice.

We compared against simdjson~\cite{simdjson}, the fastest general-purpose JSON parser, which uses SIMD to accelerate tokenization (2--6~GB/s on typical workloads). On equivalent data, Bebop decode was faster on 17 of 19 workloads (1.2--5741$\times$). The largest gaps occurred on numeric arrays: parsing ``[1.5, 2.5, ...]'' as JSON requires character-by-character float conversion, while Bebop reads IEEE 754 values directly. Simdjson was faster on JsonLarge (4.2$\times$) and JsonSmall (1.7$\times$), where JSON's native format requires no conversion.

\begin{table}[ht]
\centering
\begin{tabular}{@{}lrrr@{}}
\toprule
\textbf{Workload} & \textbf{simdjson} & \textbf{Bebop} & \textbf{Speedup} \\
\midrule
TensorShardLarge & 39.38 $\mu$s & 6.86 ns & 5741$\times$ \\
Embedding1536 & 4.69 $\mu$s & 2.80 ns & 1675$\times$ \\
EmbeddingBatch & 27.78 $\mu$s & 25.75 ns & 1079$\times$ \\
Embedding768 & 2.26 $\mu$s & 2.91 ns & 776$\times$ \\
InferenceResponse & 3.93 $\mu$s & 17.65 ns & 223$\times$ \\
OrderLarge & 283.24 ns & 5.85 ns & 48$\times$ \\
DocumentLarge & 245.53 ns & 52.0 ns & 4.7$\times$ \\
LLMChunkLarge & 1.69 $\mu$s & 677.15 ns & 2.5$\times$ \\
TreeDeep & 6.54 $\mu$s & 5.26 $\mu$s & 1.2$\times$ \\
\bottomrule
\end{tabular}
\caption{simdjson parse vs Bebop decode on equivalent data.}
\label{tab:simdjson}
\end{table}

\subsection{Roundtrip Latency}

Table~\ref{tab:roundtrip} shows encode-then-decode latency for representative workloads. Speedups range from 4.4$\times$ to 35$\times$ over Protocol Buffers.

\begin{table}[ht]
\centering
\begin{tabular}{@{}lrrrr@{}}
\toprule
\textbf{Workload} & \textbf{Protobuf} & \textbf{MsgPack} & \textbf{Bebop} & \textbf{Speedup} \\
\midrule
PersonSmall & 124.40 ns & 87.30 ns & 11.8 ns & 10.5$\times$ \\
OrderLarge & 834.68 ns & 1.01 $\mu$s & 23.86 ns & 35.0$\times$ \\
EventLarge & 238.04 ns & 150.00 ns & 52.8 ns & 4.5$\times$ \\
TreeDeep & 68.93 $\mu$s & 30.50 $\mu$s & 8.86 $\mu$s & 7.8$\times$ \\
\bottomrule
\end{tabular}
\caption{Roundtrip latency (encode + decode). Speedup is Bebop vs Protobuf.}
\label{tab:roundtrip}
\end{table}

\subsection{Wire Size}

\begin{table}[ht]
\centering
\small
\begin{tabular}{l r r r r r r}
\toprule
\textbf{Workload} & \textbf{Protobuf} & \textbf{MsgPack} & \textbf{Bebop} & \textbf{PB+brotli} & \textbf{MP+brotli} & \textbf{Bebop+brotli} \\
\midrule
\multicolumn{7}{l}{\textit{API payloads (small integer fields)}} \\
PersonSmall & 19 & 35 & 28 & --- & --- & --- \\
PersonMedium & 124 & 141 & 130 & 101 & 107 & 101 \\
OrderSmall & 34 & 94 & 76 & --- & 93 & 62 \\
OrderLarge & 423 & 664 & 1,240 & 376 & 490 & 514 \\
\midrule
\multicolumn{7}{l}{\textit{Event payloads (byte arrays)}} \\
EventSmall & 29 & 62 & 42 & --- & --- & --- \\
EventLarge & 4,158 & 4,193 & 4,170 & --- & --- & --- \\
\midrule
\multicolumn{7}{l}{\textit{ML inference (bfloat16 arrays)}} \\
Embedding768 & 1,573 & 1,568 & 1,556 & 1,193 & 1,182 & 1,167 \\
Embedding1536 & 3,109 & 3,104 & 3,092 & 2,214 & 2,246 & 2,219 \\
TensorShardSmall & 2,108 & 2,154 & 2,132 & 1,576 & 1,602 & 1,574 \\
TensorShardLarge & 65,605 & 65,652 & 65,624 & 44,720 & 44,747 & 44,711 \\
\bottomrule
\end{tabular}
\caption{Wire size in bytes. --- indicates compression increased size.}
\label{tab:wire-size}
\end{table}

The results confirm the predictions from Equation~\ref{eq:varint-expected}. OrderLarge contains arrays of 100 small integers; varint encodes these in 1--2 bytes each while fixed-width requires 4--8 bytes. MessagePack falls between the two: its schema-less design adds field name overhead, but it uses compact integer encoding.

\textbf{Compression narrows the gap on most workloads.} On ML payloads (embedding vectors, tensor shards), all three formats produce nearly identical compressed sizes---within 2\%. The bfloat16 data dominates; framing overhead becomes negligible after compression.

On API payloads with small integers (OrderLarge), compressed Protocol Buffers remains smallest (376 bytes) due to varint's compact representation of small values. MessagePack compresses to 490 bytes; Bebop to 514 bytes.

EventLarge is dominated by a 4KB random byte payload; no format benefits from compression.

With compression enabled, wire size differences between formats largely disappear. The decode performance gap does not.

\newpage
\section{Schema Language}

\subsection{Source Encoding}

Schema files (\texttt{.bop}) must be valid UTF-8. The compiler rejects files with invalid byte sequences. String literals must also be valid UTF-8; invalid sequences produce compile-time errors.

\subsection{File Structure}

\begin{minipage}{\textwidth}
\begin{minipage}[t]{0.52\textwidth}
A schema file has three sections in order:

\begin{enumerate}[leftmargin=1.5em,itemsep=2pt,topsep=4pt]
\item \textbf{Header} -- edition and package (both optional)
\item \textbf{Imports} -- import statements
\item \textbf{Definitions} -- types, constants, services
\end{enumerate}

The \texttt{edition} declaration specifies schema language version. The \texttt{package} declaration provides a namespace for all definitions.
\end{minipage}\hfill
\begin{minipage}[t]{0.44\textwidth}
\begin{lstlisting}[language=bebop,aboveskip=0pt,belowskip=0pt]
edition = "2026"
package my.app

import "bebop/decorators.bop"
import "shared/types.bop"

struct Point {
    x: float32;
    y: float32;
}
\end{lstlisting}
\end{minipage}
\end{minipage}

\subsection{Comments}

\begin{minipage}{\textwidth}
\begin{minipage}[t]{0.52\textwidth}
Three comment styles are supported.

Line comments (\texttt{//}) and block comments (\texttt{/* */}) are discarded during parsing.

\raggedright
Documentation comments (\texttt{///}) on the line immediately before a definition are captured as metadata and appear in generated code.
\end{minipage}\hfill
\begin{minipage}[t]{0.44\textwidth}
\begin{lstlisting}[language=bebop,aboveskip=0pt,belowskip=0pt]
// Line comment

/* Block comment
   spans lines */

/// Documentation comment
/// for the struct below
struct User {
    name: string;
}
\end{lstlisting}
\end{minipage}
\end{minipage}

\subsection{Literals and Escape Sequences}

String literals use double quotes (\texttt{"}) or single quotes (\texttt{'}). Literal newlines are allowed within strings.

\begin{minipage}{\textwidth}
\small
\begin{tabular}{@{}llp{2.5in}@{}}
\toprule
\textbf{Escape} & \textbf{Output} & \textbf{Notes} \\
\midrule
\texttt{\textbackslash\textbackslash} & Backslash & \\
\texttt{\textbackslash n} & Newline (LF) & \\
\texttt{\textbackslash r} & Carriage return & \\
\texttt{\textbackslash t} & Tab & \\
\texttt{\textbackslash 0} & Null byte & \\
\texttt{\textbackslash "} & Double quote & Also: \texttt{""} inside double-quoted strings \\
\texttt{\textbackslash '} & Single quote & Also: \texttt{''} inside single-quoted strings \\
\texttt{\textbackslash u\{XXXX\}} & Unicode codepoint & 1--6 hex digits, produces UTF-8 \\
\bottomrule
\end{tabular}
\end{minipage}

Numeric literals: decimal, hexadecimal (\texttt{0xFF}), scientific (\texttt{1.23e10}). Special floats: \texttt{inf}, \texttt{-inf}, \texttt{nan}. Byte arrays use a \texttt{b} prefix: \texttt{b"\textbackslash x89PNG"}.

Timestamps use ISO 8601 (\texttt{"2024-01-15T10:30:00Z"}). Timezone offsets and nanosecond precision are supported. The timezone offset can be expressed up to millisecond precision in ISO 8601-2:2019 format (\texttt{"2024-01-15T10:30:00+12:00:01.133"}).

Durations use suffixes: \texttt{"1h30m"}, \texttt{"500ms"}, \texttt{"10us"}.

String constants support environment variable substitution: \texttt{"\$(VAR)"} resolves at compile time.

\subsection{Type Reference}

See Tables~\ref{tab:primitives} and~\ref{tab:extended-types} in Section~3 for wire encoding details. Type aliases: \texttt{uint8} = \texttt{byte}, \texttt{half} = \texttt{float16}, \texttt{bf16} = \texttt{bfloat16}, \texttt{guid} = \texttt{uuid}.

\subsection{Enumerations}

\begin{minipage}{\textwidth}
\begin{minipage}[t]{0.52\textwidth}
\raggedright
Every enum must have a member with value 0 (the default).

\vspace{6pt}
Base type defaults to \texttt{uint32}. Override with a colon to specify a different underlying type.
\end{minipage}\hfill
\begin{minipage}[t]{0.44\textwidth}
\begin{lstlisting}[language=bebop,aboveskip=0pt,belowskip=0pt]
enum Status : uint8 {
    UNKNOWN = 0;
    ACTIVE = 1;
    SUSPENDED = 2;
}
\end{lstlisting}
\end{minipage}
\end{minipage}

\subsection{Structs}

\begin{minipage}[t]{0.52\textwidth}
By default, structs are immutable in generated code. Use \texttt{mut} for mutable structs. See Section~2.5 for wire encoding.
\end{minipage}\hfill
\begin{minipage}[t]{0.44\textwidth}
\begin{lstlisting}[language=bebop,aboveskip=0pt,belowskip=0pt]
struct Color {
    r: byte;
    g: byte;
    b: byte;
    a: byte;
}

mut struct MutablePoint {
    x: float32;
    y: float32;
}
\end{lstlisting}
\end{minipage}

\subsection{Messages}

\begin{minipage}[t]{0.52\textwidth}
Tags must be unique integers 1--255. Don't reuse old tags with different types. See Sections~2.5 and~2.8 for wire encoding and evolution rules.
\end{minipage}\hfill
\begin{minipage}[t]{0.44\textwidth}
\begin{lstlisting}[language=bebop,aboveskip=0pt,belowskip=0pt]
message UserProfile {
    id(1): uuid;
    name(2): string;
    email(3): string;
    created(4): timestamp;
}
\end{lstlisting}
\end{minipage}

\subsection{Unions}

\begin{minipage}[t]{0.52\textwidth}
Discriminated variant. Branches can be inline structs, inline messages, or references to existing types.
\end{minipage}\hfill
\begin{minipage}[t]{0.44\textwidth}
\begin{lstlisting}[language=bebop,aboveskip=0pt,belowskip=0pt]
union Result {
    Success(1): {
        value: string;
    };
    Error(2): {
        code: int32;
        message: string;
    };
}
\end{lstlisting}
\end{minipage}

\subsection{Services}

\begin{minipage}{\textwidth}
\begin{minipage}[t]{0.48\textwidth}
\raggedright
Services define RPC interfaces. The \texttt{stream} keyword indicates multiple messages. Request and response types must be named struct, message, or union definitions. Primitives and inline types are not allowed.

Services can include methods from other services using the \texttt{with} keyword for composition.
\end{minipage}\hfill
\begin{minipage}[t]{0.48\textwidth}
\begin{lstlisting}[language=bebop,aboveskip=0pt,belowskip=0pt]
service BaseService {
    GetStatus(Req): StatusRes;
}

service ChatService with BaseService {
    Send(Message): Ack;
    Subscribe(Req): stream Event;
    Upload(stream Chunk): Summary;
    Chat(stream Msg): stream Msg;
}
\end{lstlisting}
\end{minipage}
\end{minipage}

\begin{minipage}{\textwidth}
\small
\begin{tabular}{@{}llp{2in}@{}}
\toprule
\textbf{Type} & \textbf{Syntax} & \textbf{Description} \\
\midrule
Unary & \texttt{Method(Req): Res} & Single request, single response \\
Server stream & \texttt{Method(Req): stream Res} & Request, multiple responses \\
Client stream & \texttt{Method(stream Req): Res} & Multiple requests, response \\
Bidirectional & \texttt{Method(stream): stream} & Streams both directions \\
\bottomrule
\end{tabular}
\end{minipage}

\begin{minipage}{\textwidth}
\subsection{Constants}

\begin{minipage}[t]{0.48\textwidth}
Named compile-time values accessible in generated code.

Timestamp literals use ISO 8601 format. Duration literals use time unit suffixes: \texttt{h}, \texttt{m}, \texttt{s}, \texttt{ms}, \texttt{us}, \texttt{ns}.

Byte arrays use a \texttt{b} prefix with escape sequences.
\end{minipage}\hfill
\begin{minipage}[t]{0.48\textwidth}
\begin{lstlisting}[language=bebop,aboveskip=0pt,belowskip=0pt]
const int32 MAX_SIZE = 1024;
const string HOST = "localhost";
const timestamp EPOCH =
    "1970-01-01T00:00:00Z";
const duration TIMEOUT = "30s";
const byte[] PNG_MAGIC =
    b"\x89PNG\r\n\x1a\n";
\end{lstlisting}
\end{minipage}

\vspace{12pt}
\subsection{Visibility}

\begin{minipage}[t]{0.52\textwidth}
\raggedright
Top-level definitions are exported by default. Use \texttt{local} to make them file-private.

Nested definitions (types inside structs, messages, or unions) are local by default. Use \texttt{export} to make them accessible.
\end{minipage}\hfill
\begin{minipage}[t]{0.44\textwidth}
\begin{lstlisting}[language=bebop,aboveskip=0pt,belowskip=0pt]
struct PublicType {}
local struct PrivateType {}

struct Outer {
    struct LocalInner {}
    export struct PublicInner {}
}
\end{lstlisting}
\end{minipage}
\end{minipage}

\subsection{Compile-Time Extensibility}

\begin{minipage}{\textwidth}
Decorators provide schema annotations processed at compile time. Code generators receive decorator arguments through the standard plugin protocol. The \texttt{export} block extends this by computing derived values from the full schema context.

\vspace{4pt}
\begin{minipage}[t]{0.44\textwidth}
\small
\textbf{Decorator syntax:}
\begin{itemize}[leftmargin=1.2em,itemsep=0pt,topsep=2pt]
\item \texttt{targets} -- where decorator applies
\item \texttt{param name!: Type} -- required
\item \texttt{param name?: Type} -- optional
\item \texttt{validate [[ lua ]]} -- reject invalid usage
\item \texttt{export [[ lua ]]} -- produce plugin metadata
\end{itemize}

\vspace{2pt}
{\footnotesize Valid targets: \texttt{ENUM}, \texttt{STRUCT}, \texttt{MESSAGE}, \texttt{UNION}, \texttt{FIELD}, \texttt{SERVICE}, \texttt{METHOD}, \texttt{BRANCH}, \texttt{ALL}.}

\vspace{4pt}
The \texttt{validate} block receives parameters plus a \texttt{target} table (kind, name, parent).
\end{minipage}\hfill
\begin{minipage}[t]{0.52\textwidth}
\begin{lstlisting}[language=bebop,aboveskip=0pt,belowskip=0pt,basicstyle=\ttfamily\footnotesize]
#decorator(range) {
    targets = FIELD
    param min!: int32
    param max!: int32
    validate [[
        if min >= max then
            error("min must be < max", self.min.span)
        end
    ]]
    export [[
        return { range_min = min, range_max = max,
                 width = max - min }
    ]]
}
\end{lstlisting}
\end{minipage}

\vspace{6pt}
\textbf{Export blocks} compute derived data using the full schema context. Plugins already receive raw decorator arguments through the descriptor. The export block produces values that require compile-time computation or access to the target element's metadata.

\vspace{4pt}
\begin{minipage}[t]{0.58\textwidth}
\begin{lstlisting}[language=bebop,basicstyle=\ttfamily\footnotesize]
#decorator(indexed) {
    targets = FIELD
    param unique?: bool
    export [[
        local t, f = target.parent, target.name
        return {
            index_name = t.."_"..f.."_idx",
            table_name = t, column_name = f,
            is_unique = unique or false
        }
    ]]
}
\end{lstlisting}
\end{minipage}\hfill
\begin{minipage}[t]{0.38\textwidth}
\small
\raggedright
A database code generator receives the computed \texttt{index\_name} directly (e.g., \texttt{"User\_email\_idx"}).

\vspace{4pt}
Export blocks can aggregate data across the schema for deprecation reports, validation constraints, or reflection tables.
\end{minipage}
\end{minipage}

\subsection{Schema Evolution}
\label{sec:evolution}

\begin{table}[ht]
\centering
\small
\begin{tabular}{@{}llcl@{}}
\toprule
\textbf{Type} & \textbf{Change} & \textbf{Safe?} & \textbf{Notes} \\
\midrule
\multirow{5}{*}{Message}
  & Add field & \cmark & Use new tag; old readers ignore \\
  & Deprecate field & \cmark & Skipped on wire; don't reuse tag \\
  & Rename field & \cmark & Names not on wire \\
  & Change field type & \xmark & Never reuse tag with different type \\
  & Change tag number & \xmark & Equivalent to remove + add \\
\midrule
\multirow{4}{*}{Struct}
  & Add field & \xmark & Positional encoding; no tags \\
  & Remove field & \xmark & Create versioned type instead \\
  & Reorder fields & \xmark & Or convert to message \\
  & Change field type & \xmark & \\
\midrule
\multirow{3}{*}{Union}
  & Add branch & \cmark & \\
  & Remove branch & \xmark & Decode fails for existing data \\
  & Change branch type & \xmark & \\
\midrule
\multirow{3}{*}{Enum}
  & Add value & \cmark & \\
  & Remove value & \xmark & Existing data may contain it \\
  & Change base type & \xmark & e.g., \texttt{uint8} to \texttt{uint32} \\
\bottomrule
\end{tabular}
\caption{Schema evolution compatibility. \cmark\ = backward compatible, \xmark\ = breaking change.}
\end{table}

\newpage
\section{Implementation}

\subsection{Compiler}

The Bebop compiler (\texttt{bebopc}) is written in portable C with no external dependencies. The resulting binary is under 2MB and runs on Linux, macOS, Windows, and BSD.

\begin{lstlisting}[language={}]
$ bebopc build schema.bop --c_out=./generated
$ bebopc build schema.bop --typescript_out=./ts
\end{lstlisting}

\subsection{Plugin Architecture}

Code generators are standalone executables named \texttt{bebopc-gen-\$NAME}. The compiler discovers them in PATH and invokes them via \texttt{--\$NAME\_out=DIR}.

\begin{minipage}{\textwidth}
\begin{minipage}[t]{0.48\textwidth}
\raggedright
Communication uses Bebop-encoded messages on stdin/stdout. Plugins can be written in any language with a Bebop runtime.

\textbf{Request fields:}
\begin{itemize}[leftmargin=1.2em,itemsep=1pt,topsep=2pt]
\item \texttt{files\_to\_generate} -- source files from command line
\item \texttt{parameter} -- from \texttt{--\$NAME\_opt=...}
\item \texttt{schemas} -- descriptors, topologically sorted
\end{itemize}

\textbf{Response fields:}
\begin{itemize}[leftmargin=1.2em,itemsep=1pt,topsep=2pt]
\item \texttt{error} -- fatal error message (if any)
\item \texttt{files} -- generated file name + content pairs
\item \texttt{diagnostics} -- warnings/errors with source spans
\end{itemize}
\end{minipage}\hfill
\begin{minipage}[t]{0.48\textwidth}
\begin{lstlisting}[language=bebop,aboveskip=0pt,belowskip=0pt]
message CodeGeneratorRequest {
    files_to_generate(1): string[];
    parameter(2): string;
    compiler_version(3): Version;
    schemas(4): SchemaDescriptor[];
}

message CodeGeneratorResponse {
    error(1): string;
    files(2): GeneratedFile[];
    diagnostics(3): Diagnostic[];
}
\end{lstlisting}
\end{minipage}
\end{minipage}

Plugins can extend files from other plugins using insertion points, markers that later plugins can target.

\subsection{Descriptor Format}

The compiled schema representation uses Bebop's own wire format. Descriptors are passed to plugins and can be used for runtime reflection.

\begin{minipage}{\textwidth}
\begin{minipage}[t]{0.52\textwidth}
\raggedright
\textbf{Structure:}
\begin{itemize}[leftmargin=1.2em,itemsep=1pt,topsep=2pt]
\item \texttt{DescriptorSet} -- root container
\item \texttt{SchemaDescriptor[]} -- one per .bop file
\item \texttt{DefinitionDescriptor[]} -- types, services, constants
\end{itemize}

Definitions are topologically sorted: dependencies appear before dependents. Process sequentially for single-pass code generation.

Each definition includes its kind (enum, struct, message, union, service, const), fully-qualified name, documentation from \texttt{///} comments, visibility, and applied decorators with their exported data.
\end{minipage}\hfill
\begin{minipage}[t]{0.44\textwidth}
\begin{lstlisting}[language=bebop,aboveskip=0pt,belowskip=0pt]
message DefinitionDescriptor {
    kind(1): DefinitionKind;
    name(2): string;
    fqn(3): string;
    documentation(4): string;
    visibility(5): Visibility;
    decorators(6): DecoratorUsage[];
    nested(7): DefinitionDescriptor[];
    // kind-specific body:
    enum_def(8): EnumDef;
    struct_def(9): StructDef;
    message_def(10): MessageDef;
    union_def(11): UnionDef;
    service_def(12): ServiceDef;
    const_def(13): ConstDef;
}
\end{lstlisting}
\end{minipage}
\end{minipage}

Type references use \texttt{TypeDescriptor} with a \texttt{kind} field (BOOL, INT32, STRING, ARRAY, MAP, DEFINED, etc.) and recursive structure for nested types. Service methods include a stable 32-bit routing ID computed from \texttt{/ServiceName/MethodName} using MurmurHash3 with the \texttt{lowbias32} finalizer~\cite{hashprospector} (bias 0.17 vs.\ standard 0.23).

\newpage
\section{RPC Protocol}

\subsection{Motivation}

In microservice architectures, serialization sits inside the RPC layer. The framing, error encoding, metadata handling, and call multiplexing all add latency on top of the serialization itself. gRPC couples Protocol Buffers to HTTP/2 with its own framing protocol, length-prefixed messages, and HPACK-compressed headers~\cite{grpc-http2-spec}. Replacing the serialization format means replacing the RPC stack.

Bebop's RPC protocol uses Bebop encoding for every layer: frame headers, call headers, error payloads, metadata, the batch protocol, and service discovery responses. An implementation that can decode Bebop messages can decode every part of the protocol. One encoding, one set of generated types, one decoder path.

\subsection{Protocol Design}

A Bebop RPC frame has a fixed 9-byte header:

\vspace{4pt}
\begin{center}
\begin{tikzpicture}
\wirefield{0}{3}{length}
\wirefield{3}{4}{flags}
\wirefield{4}{8}{stream\_id}
\wirelabel{0}{3}{uint32}
\wirelabel{3}{4}{byte}
\wirelabel{4}{8}{uint32}
\end{tikzpicture}
\end{center}
\vspace{4pt}

\noindent The \texttt{length} field is the payload byte count. \texttt{flags} is a bitfield combining \texttt{END\_STREAM}, \texttt{ERROR}, \texttt{COMPRESSED}, \texttt{TRAILER}, and \texttt{CURSOR}. \texttt{stream\_id} provides multiplexing on transports that require it. A complete unary RPC uses 18 bytes of framing overhead: 9 bytes in each direction.

The protocol supports four method types: unary (single request, single response), server streaming (one request, multiple responses), client streaming (multiple requests, one response), and duplex streaming (both directions). These match gRPC's taxonomy, and status codes 0--16 align with gRPC's definitions, so bridging between the two protocols requires no code remapping.

Method dispatch uses a 4-byte hash of \texttt{/ServiceName/MethodName} computed at schema compile time via MurmurHash3. The router performs integer comparison instead of string matching on every incoming call.

The protocol is transport-agnostic. On HTTP/1.1 and HTTP/2, each request-response pair maps to a standard HTTP exchange, with metadata carried in HTTP headers. On binary transports (TCP, WebSocket, IPC, Unix sockets), the full frame protocol runs directly, with a \texttt{CallHeader} initiating each call and stream IDs providing multiplexing. This separation means the same service definition generates handlers that deploy identically on an HTTP load balancer, a WebSocket gateway, or a raw TCP socket.

\subsection{Batch Pipelining}
\label{sec:batch-pipelining}

Dependent cross-service calls typically require sequential round trips. Fetching a user record and then querying their friends list costs two round trips: the second call cannot begin until the first completes. In a chain of $N$ dependent calls, latency grows as $N \times \text{RTT}$.

Bebop RPC provides a batch protocol that collapses dependent calls into a single round trip. Each call in a batch carries an \texttt{input\_from} field. When set to $-1$, the call uses its own payload. When set to the index of a previous call, the server forwards that call's result as input. The server builds a dependency graph from these references, partitions calls into execution layers, and runs all calls within a layer concurrently. Layer $k+1$ waits only for the calls in layer $k$ that it depends on.

\begin{lstlisting}[language=bebop]
message BatchCall {
    call_id(1): int32;
    method_id(2): uint32;
    payload(3): byte[];
    input_from(4): int32;  // -1 = use payload, >=0 = forward result
}
\end{lstlisting}

If a call fails, all calls that depend on it also fail with status \texttt{INVALID\_ARGUMENT}. If the batch deadline expires mid-execution, remaining calls fail with \texttt{DEADLINE\_EXCEEDED}. Server-stream methods within a batch buffer their results into arrays; true streaming semantics would require multiplexed response framing, so client-stream and duplex methods are excluded from batching.

gRPC has no built-in batching primitive~\cite{grpc-core-concepts}. Dependent calls require sequential round trips or application-level batch message types defined in the Protobuf schema. Bebop's batch protocol handles dependency resolution and concurrent execution at the framework level.

\subsection{Deadline Propagation}

Bebop RPC transmits deadlines as absolute timestamps with nanosecond precision. Every downstream hop checks the same wall-clock cutoff. The Google SRE book identifies deadline propagation as a defense against cascading failures~\cite{google-sre-cascading}. Without propagation, downstream services continue working on requests the caller has already abandoned.

gRPC converts absolute deadlines to relative timeouts with elapsed time deducted at each hop~\cite{grpc-deadlines-guide}. Bebop avoids the deduction step by transmitting the absolute timestamp directly on binary transports. On HTTP transports, the deadline is a millisecond Unix timestamp in the \texttt{bebop-deadline} header. Both representations refer to the same wall-clock instant, so no rounding accumulates across hops.

\subsection{Stream Cursors}
\label{sec:stream-cursors}

A server-stream call that delivers 10{,}000 results and drops at result 9{,}500 has two options without protocol support: re-request the entire stream, or track progress in the application layer. gRPC takes the second approach---the protocol has no resumption mechanism, so handlers implement their own checkpointing.

Bebop RPC provides cursor-based resumption at the frame level. The \texttt{CallHeader} carries a \texttt{cursor} field (uint64). On the first call this is zero. On reconnection, the client sends the last cursor it fully processed. The handler reads \texttt{RpcContext.cursor} and skips past already-delivered data. In the other direction, response frames carry position markers: when the \texttt{CURSOR} flag (0x10) is set, 8 bytes of little-endian uint64 follow the payload. The \texttt{length} field counts only payload bytes; the cursor is appended outside it. Not every frame needs a cursor, and a stream may freely mix cursored and non-cursored frames.

What a cursor value means is handler-specific: a database offset, a sequence number, a timestamp, a log position. The protocol treats it as an opaque uint64. Resume logic stays in the call header and wire frames, never in application message types. This separation is also why only unary methods can be dispatched as futures (Section~\ref{sec:futures})---server-stream methods already have a protocol-level mechanism for surviving disconnections.

\subsection{Futures}
\label{sec:futures}

Long-running operations create a tension in synchronous RPC. An ML inference call that takes 30 seconds holds a connection and a server thread for the duration. The caller cannot disconnect and reconnect without losing the result. gRPC's standard pattern for this is Google's \texttt{google.longrunning.Operations} service~\cite{google-lro}, which returns an operation ID and requires the client to poll a separate \texttt{GetOperation} endpoint until completion.

Bebop RPC replaces polling with push-based delivery using three reserved method IDs:

\begin{table}[H]
\centering
\small
\begin{tabular}{@{}clll@{}}
\toprule
\textbf{ID} & \textbf{Method} & \textbf{Type} & \textbf{Payload} \\
\midrule
2 & Dispatch & unary & \texttt{FutureDispatchRequest} $\to$ \texttt{FutureHandle} \\
3 & Resolve & server-stream & \texttt{FutureResolveRequest} $\to$ \texttt{FutureResult} \\
4 & Cancel & unary & \texttt{FutureCancelRequest} $\to$ \texttt{Empty} \\
\bottomrule
\end{tabular}
\end{table}

\noindent A \texttt{FutureDispatchRequest} wraps a unary call or batch for background execution. The server registers the work, spawns a task, and returns a \texttt{FutureHandle} containing a server-generated v4 UUID. The dispatch call completes in the time it takes to validate the request and allocate the ID; the \texttt{deadline} field in the request applies to the inner call's execution, not to the dispatch itself.

The resolve stream (method~3) is a server-stream connection that pushes \texttt{FutureResult} messages as futures complete. Each result contains the future's UUID and a terminal outcome: either a success payload with response metadata, or an error with a status code. The \texttt{ids} field in the request filters delivery to specific futures; omitting it subscribes to all futures owned by the caller. When specific IDs are requested and some have already completed, the server sends those results immediately before continuing with remaining completions. The inner handler is unaware it is running as a future---the server invokes it identically to a synchronous unary call.

\subsubsection{Idempotency and ownership}

The \texttt{FutureDispatchRequest} carries an optional \texttt{idempotency\_key} field, a client-generated UUID. If a pending or completed future with the same key exists for the same caller, the server returns the existing handle without dispatching again. Cancellation releases the key so a subsequent dispatch creates a new future. Keys are scoped per caller; two different callers can use the same key without collision.

Every future is bound to a caller identity resolved from peer information: authenticated identity if available, otherwise the connection's remote address. The server checks this identity on every resolve and cancel operation. A caller that does not own a future receives \texttt{PERMISSION\_DENIED}.

\subsubsection{Retention and storage}

The server configures a default retention policy---typically eviction-by-count---that applies to all completed futures. Clients can override this per-dispatch by setting \texttt{discard\_result} in the dispatch request. When set, the server delivers the result to active resolve streams and immediately discards it---the future is not promised. The handle is still returned so the client can cancel in-flight work, and idempotency keys still deduplicate retries, but rehydration from a saved UUID returns nothing because the result no longer exists. Clients cannot force retention beyond the server's policy; they can only opt out of it.

The reference implementation defines an asynchronous storage protocol so that the in-memory store can be replaced with database, disk, or tiered-cache backends. Protocol operations are split for composability: persisting a completed result and notifying active subscribers are separate interface methods, allowing a database backend to commit before fanning out to in-memory streams.

\subsection{Deployment Model}

gRPC requires HTTP/2~\cite{grpc-http2-spec}. This creates friction in several environments. AWS Lambda's API Gateway communicates with functions over HTTP/1.1; a Coinbase engineering investigation found gRPC ``very close to possible'' for unary calls but hit blocking issues with trailing headers, and streaming was unsupported entirely~\cite{henry-grpc-lambda}. Cloudflare Workers lack HTTP/2 streaming support in the \texttt{workerd} runtime~\cite{cloudflare-workerd-grpc}. In browsers, gRPC-Web requires an Envoy proxy because browser APIs do not expose HTTP/2 framing or trailing headers~\cite{brandhorst-grpc-web}.

Twitch built Twirp~\cite{nelson-twirp} as an HTTP/1.1-compatible alternative to gRPC after grpc-go's embedded HTTP/2 implementation caused production outages. Twirp drops streaming entirely in exchange for deployment simplicity.

Bebop RPC is transport-agnostic. Unary and streaming methods work over HTTP/1.1 through HTTP/3, binary TCP, WebSocket, WebRTC, WebTransport, and IPC. In browsers, any available transport works directly with no proxy. Metadata maps to HTTP headers. Errors map to HTTP status codes. The same \texttt{.bop} service definition generates server handlers and typed client stubs for every target language.

\subsection{Comparison with gRPC}

\begin{table}[H]
\centering
\small
\begin{tabular}{@{}lll@{}}
\toprule
\textbf{Feature} & \textbf{gRPC} & \textbf{Bebop RPC} \\
\midrule
Wire encoding & Protobuf + HTTP/2 framing & Bebop throughout \\
Frame overhead & 5B length prefix + HTTP/2 frame & 9B fixed header \\
Batch pipelining & Application-level & Native with dependency graph \\
Async dispatch & Polling (\texttt{google.longrunning}) & Push-based futures \\
Transport & HTTP/2 required & Any \\
Serverless & Limited & HTTP/1.1 compatible \\
Method dispatch & String path matching & 4-byte hash comparison \\
Deadline model & Relative timeout with deduction & Absolute timestamp \\
Status codes & 0--16 & 0--16 aligned + 17--255 app-defined \\
Browser support & Requires Envoy proxy & Native \\
\bottomrule
\end{tabular}
\caption{Feature comparison between gRPC and Bebop RPC.}
\label{tab:rpc-comparison}
\end{table}

\section{Related Work}

Protocol Buffers~\cite{protobuf} prioritizes wire compactness through varint encoding. The branching cost during decode was recognized early~\cite{varda-hn}. Bebop prioritizes decode speed instead.

Cap'n Proto~\cite{capnproto} and FlatBuffers~\cite{flatbuffers} optimize for random field access through pointer-based layouts. Bebop targets sequential decode, which matches how ML pipelines consume embedding vectors.

Simple Binary Encoding (SBE)~\cite{sbe} also uses fixed-width encoding, targeting financial systems where latency matters. SBE focuses on FIX protocol semantics and requires fields in schema-defined order. Bebop takes a similar approach to wire encoding but provides a more general-purpose schema language with tagged messages for evolution.

MessagePack~\cite{msgpack} embeds type tags in the wire format; Avro~\cite{avro} includes the schema with each message. Both add per-message overhead that Bebop avoids through code generation.

simdjson~\cite{simdjson} uses SIMD to accelerate JSON parsing. Section~\ref{sec:eval} compares performance.

Bebop's RPC protocol builds on ideas from several frameworks. gRPC~\cite{grpc-http2-spec} provides RPC over HTTP/2 with Protocol Buffers as the default wire format. Bebop aligns its status codes with gRPC's and adopts the same four method types, but removes the HTTP/2 requirement and adds batch pipelining. Section~7 details these differences.

Apache Thrift~\cite{slee-thrift} combines an IDL, multiple serialization formats, and transport abstraction in a single framework. Thrift predates gRPC and influenced its design.

Twirp~\cite{nelson-twirp} dropped streaming entirely in favor of HTTP/1.1 compatibility and deployment simplicity. Bebop RPC retains streaming by supporting multiple transports.

Google's long-running operations pattern~\cite{google-lro} adds async dispatch to gRPC through a polling-based \texttt{Operations} service. The client calls \texttt{GetOperation} until the \texttt{done} field is true. Bebop's futures use a push-based resolve stream instead, avoiding repeated round trips but requiring the client to maintain a stream connection.

\section{Conclusion}

Varint encoding saved bytes at the cost of branch-heavy decode loops. As the gap between bandwidth growth and compute growth widens, that tradeoff ages poorly.

Bebop decodes tensor data 213$\times$ faster than Protocol Buffers. Fixed-width encoding eliminates the branch-per-byte loop that varints require. No tag dispatch for structs. No runtime length computation. The decoder reduces to pointer arithmetic and bounds checks, simple enough that it saturates memory bandwidth on large records.

The tradeoff is wire size. Fixed-width integers cost more bytes than varints for small values. Section~\ref{sec:eval} shows where this matters and where compression eliminates the difference.

Faster serialization matters more when the RPC layer preserves the gain. Batch pipelining turns $N$ dependent calls into a single exchange, futures replace polling for long-running operations with push-based delivery, and HTTP/1.1 compatibility means the protocol deploys where gRPC cannot.

What surprised us was how much performance came from doing less. Fixed sizes, predictable layouts, no per-value decisions. The same choices that make Bebop fast also keep the compiler small. The implementation is 35,000 lines of C with single-pass code generation. Simplicity compounded.

Future work includes SIMD-accelerated array decoding, GPU-side deserialization for direct device memory placement, streaming decode for records larger than available memory, and RPC runtime implementations for additional target languages. The wire format and RPC protocol are stable; these are implementation improvements that maintain compatibility.

\section*{Acknowledgments}

We thank Tristram Jenkins, PhD, of Tokyo University and Yoonseo Kang for detailed feedback on earlier drafts of this paper.

\section*{Availability}

Bebop is open source. The compiler, runtime libraries, and benchmark code are available at \url{https://github.com/6over3}.

\end{document}